\documentclass[aip,amsmath,amssymb,reprint]{revtex4-2}
\usepackage{xurl} % So that URLs in the references break/wrap.
\usepackage{hyperref} % Clickable links.
\usepackage{graphicx} % Figures.

\usepackage[utf8]{inputenc}
\usepackage[T1]{fontenc}
\usepackage{mathptmx}
\usepackage{etoolbox}

\begin{document}

%-------------------------------------------------------------------------------
% Front Matter
%-------------------------------------------------------------------------------

\title{Improving Student Self-Confidence in Quantum Computing with the\\Qubit Touchdown Board Game}

\author{Kristina Armbruster}
    \email{kristina.armbruster@bpsne.net}
    \affiliation{Bellevue West High School, 1501 Thurston Ave, Bellevue, Nebraska 68123}
\author{Gintaras Duda}
    \email{gkduda@creighton.edu}
    \affiliation{Department of Physics, Creighton University, 2500 California Plaza, Omaha, Nebraska 68178}
\author{Thomas G. Wong}
    \email{thomaswong@creighton.edu}
    \affiliation{Department of Physics, Creighton University, 2500 California Plaza, Omaha, Nebraska 68178}

\maketitle

%-------------------------------------------------------------------------------
% Main Matter
%-------------------------------------------------------------------------------

Qubit Touchdown is a two-player, competitive board game that was developed to introduce students to quantum computing. A quantum computer is a new kind of computer that is based on the laws of quantum physics, and it can solve certain problems faster than normal computers because it follows a different set of rules. Qubit Touchdown's game play mirrors the rules of (American) football, with players taking turns moving the football to score the most touchdowns, and no knowledge of quantum computing is needed to play the game. We evaluated the game with 107 public high school students in Precalculus, Advanced Placement (AP) Statistics, and/or AP Physics 1 courses, assessing whether their interest in and self-confidence in their ability to learn  quantum computing changed as a result of playing the game and learning about its connections to quantum computing. We also assessed whether the game was easy to learn and enjoyable. We found that students' interest in quantum computing increased slightly ($p<0.05$), but students' self-confidence in their ability to learn quantum computing saw greater gains ($p<0.001$); students also widely considered the game accessible and fun. Thus, Qubit Touchdown could be an effective resource to introduce students to Quantum Computing and boost their confidence in learning about the field. Free printables of the game are available, and professionally produced copies can be purchased on demand.

%-------------------------------------------------------------------------------
% Section
%-------------------------------------------------------------------------------

\section{Quantum Games}

Quantum computing is a significant component of a larger field called quantum information science (QIS), where various properties of quantum mechanics are exploited for information gathering, processing, and sharing. The U.S. government has identified quantum technologies as ``critical and emerging technologies,'' which are ``advanced technologies that are potentially significant to U.S.~national security.'' \cite{CETs2024} As part of whole-of-government efforts enabled by the National Quantum Initiative\cite{NQIA,Raymer2019} to ensure continued U.S. leadership in the field, the National Science and Technology Council (NSTC) Subcommittee on QIS published a national strategic plan on developing a quantum information science and technology (QIST) workforce,\cite{QIST-Workforce-Strategy} and it explicitly calls out the need for learners to be ``provided exposure to QIST via accessible outreach and educational opportunities [including] in informal learning venues such as museums, movies, games, and other media.'' Note that games are listed, and the report further recommends that such games and activities begin with K-12 age groups.

\begin{figure}
    \includegraphics[width=\linewidth]{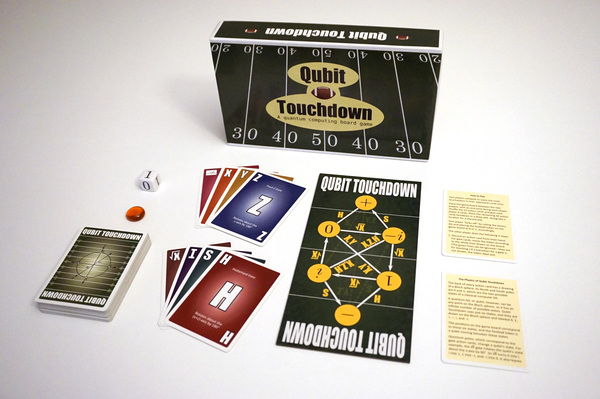}
    \caption{\label{fig:overview}Qubit Touchdown board game.}
\end{figure}

Games on various platforms have been created to address this need. Computer games may mimic quantum phenomena as a way of building intuition about quantum mechanics, such Quantum Tic-Tac-Toe,\cite{QuantumTicTacToeCompadre,Goff2004} the qCraft\cite{qCraft} extension to Minecraft, and Quantum Chess.\cite{QuantumChess} QiskitBlocks\cite{QiskitBlocks} is a game for the Minecraft-like game engine Luanti, formerly known as Minetest, where players learn to program quantum circuits by going from one escape room to the next. Hackathons have also been a source of computer games to teach quantum concepts, such as QPong,\cite{QPong} a quantum version of the classic arcade game Pong, and numerous outputs of Quantum Game Jams in Finland. \cite{QuantumWheel} Some computer games partially run on actual quantum processors, such as Cat-Box-Scissors \cite{Wootton2017} and Quantum Battleship.\cite{Wootton2018} Games have also been made for mobile devices, such as Mequanic,\cite{Tahan2015,Mequanic} Hello Quantum,\cite{Wootton2018} Quantum Cats,\cite{QuantumCats} and Tiq Taq Toe\cite{TiqTaqToe} (which has different rules from Quantum Tic-Tac-Toe). Finally, some tabletop games exist, such as Entanglion\cite{Entanglion} and Brakets,\cite{Brakets} as well as the subject of this article, Qubit Touchdown,\cite{QubitTouchdown-Overview} shown in Fig.~\ref{fig:overview}.

%-------------------------------------------------------------------------------
% Section
%-------------------------------------------------------------------------------

\section{\label{sec:howtoplay}How to Play Qubit Touchdown}

\begin{figure}
    \includegraphics[width=1.75in]{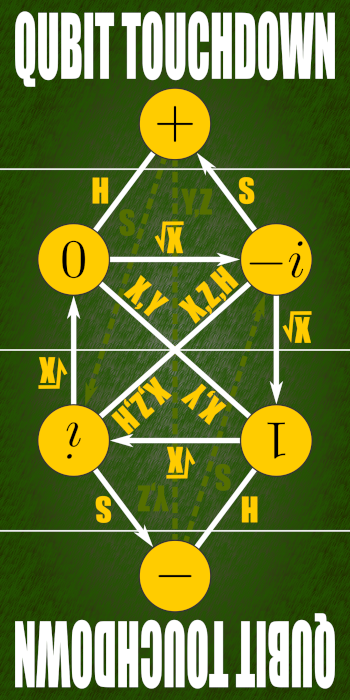}
    \caption{\label{fig:board}Qubit Touchdown board.}
\end{figure}

As a high-level overview, Qubit Touchdown is a two player, competitive game. The game board is shown in Fig.~\ref{fig:board}, and the football token or gem is placed on the board to indicate the position of the football. The football can only be at the six positions indicated by yellow circles, which are labeled $0$, $1$, $+$, $-$, $i$, and $-i$. One player is trying to get the football to the $+$ endzone to score touchdowns, and the other player is trying to get the ball to the $-$ endzone to score touchdowns. They take turns playing cards to move the football along the white lines/arrows on the board. Whoever scores the most touchdowns by the time all 52 cards are played wins.

\begin{figure}
    \includegraphics[width=3in]{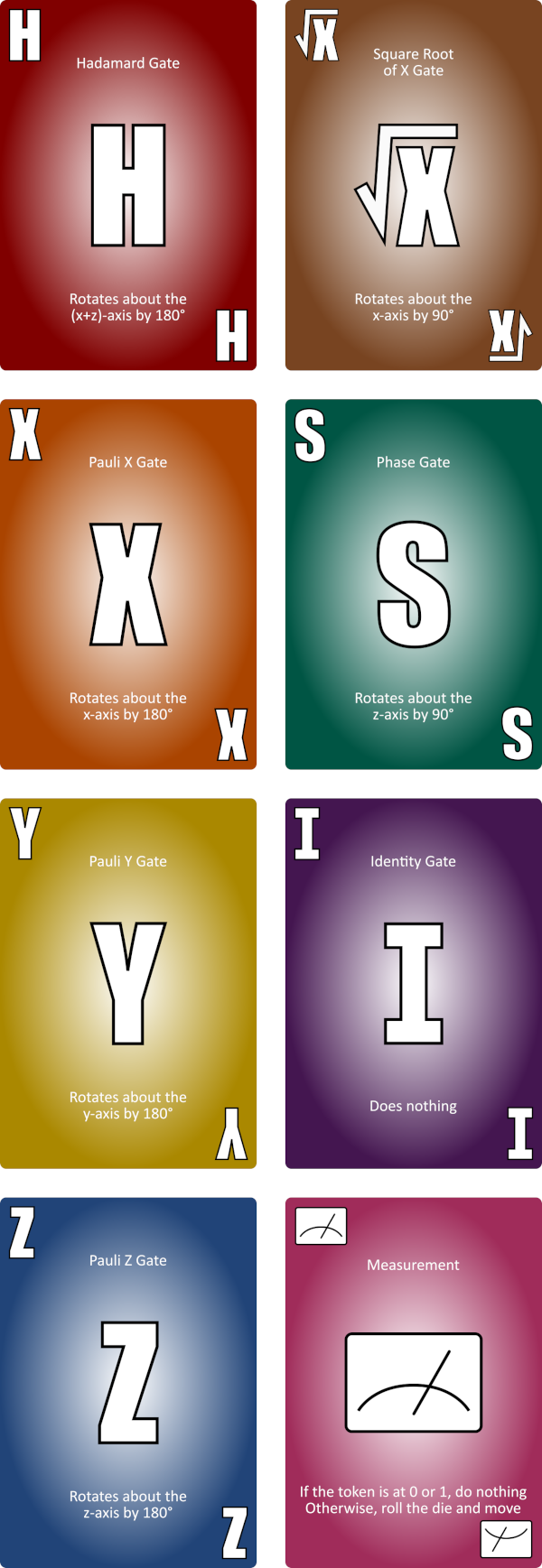}
    \caption{\label{fig:cards}Qubit Touchdown cards.}
\end{figure}

To elaborate, the game is set up by shuffling all 52 cards and distributing four to each player. The remaining 44 cards are placed in a ``draw pile.'' A player ``kicks off'' by rolling a binary die, whose only outcomes are 0 or 1, each with a 50:50 chance. They place the football at position $0$ or $1$, matching the outcome of the roll. The other player ``returns'' the football by playing an action card. The cards are $X$, $Y$, $Z$, $I$, $H$, $S$, $\sqrt{X}$, and Measurement, as shown in Fig.~\ref{fig:cards}. For all of the single-lettered cards, the movements are indicated on the board using white lines or arrows, and if a transition does not appear, the football stays put. For example, at position $0$, playing $H$ moves the football to $+$, $\sqrt{X}$ moves it to $-i$, $X$ or $Y$ moves it to $1$, and $S$, $Z$, and $I$ do not have a line from position $0$, so the football stays at $0$ if any of these are played. Note that lines are bi-directional, so $X$ would move the football from $1$ back to $0$, but the arrows are uni-directional, so $\sqrt{X}$ would move the football from $-i$ to $1$, not back to $0$. The measurement card is like attempting to punt the football: if the football is at position $0$ or $1$, nothing happens (a failed punt), but if it is anywhere else, the player rolls the binary die and places the ball at $0$ or $1$ according to the outcome of the roll (a successful punt, although depending on the outcome, it may not be a very good punt). After playing a card, the player draws another one. If they scored a touchdown, they ``kick off,'' and now it is the other player's turn to play a card. Players alternate playing and replenishing cards, so they always have four cards in their hand, until the end of the game when the draw pile is depleted, and then they continue taking turns playing whatever cards they have until they have none left. Whoever scored the most touchdowns wins.

%-------------------------------------------------------------------------------
% Section
%-------------------------------------------------------------------------------

\section{\label{sec:connections}Connections to Quantum Computing}

\begin{figure}
    \includegraphics[width=2in]{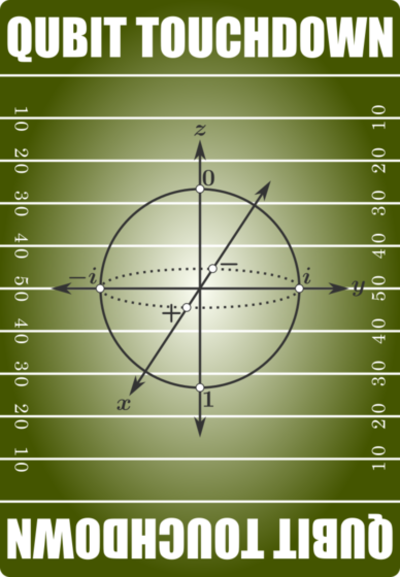}
    \caption{\label{fig:blochsphere}The back of each card is a Bloch sphere.}
\end{figure}

Qubit Touchdown's connections to quantum computing can be explained using the back of any game card, which is shown in Fig.~\ref{fig:blochsphere}. It contains a drawing of a sphere called a \emph{Bloch sphere}, with the $x$-axis coming out of the card, the $y$-axis to the right, and the $z$-axis up. The north pole, which lies along the $z$-axis, is 0, and the south pole, which lies along the $-z$\nobreakdash-axis, is 1. A traditional computer bit can only be in the $0$ or $1$ state, so it can only be at the north pole or the south pole. A quantum bit, or \emph{qubit}, however, can be at any point on the Bloch sphere. Besides being at the north pole or the south pole, it can be anywhere in the northern hemisphere, southern hemisphere, or equator. Qubit Touchdown uses just six possible locations: $0$, $1$, $+$, $-$, $i$, and $-i$, which respectively lie on the $z$-axis, $-z$-axis, $x$-axis, $-x$-axis, $y$-axis, and $-y$-axis. The positions on the game board in Fig.~\ref{fig:board} correspond to these six locations, and the football is a qubit moving between these six states.

With the exception of the measurement card, the rest of the Qubit Touchdown cards correspond to \emph{quantum gates}, which change the qubit's state. The description of each quantum gate is listed on the card, as shown in Fig.~\ref{fig:cards}. For example, the $X$ card is a Pauli $X$ Gate, which rotates the qubit's state on the Bloch sphere about the $x$-axis by $180^\circ$. Looking at the Bloch sphere in Fig.~\ref{fig:blochsphere}, if a qubit is at $0$ at the north pole, then rotating it by $180^\circ$ about the $x$-axis would move it to $1$ at the south pole. If it were rotated again by $180^\circ$ around the $x$-axis, it would return to $0$. This is why $X$ card moves the football between positions $0$ and $1$ on the game board in Fig.~\ref{fig:board}. All quantum gates on a qubit rotate its state on the Bloch sphere by some angle around some axis, and the cards in Fig.~\ref{fig:cards} indicate the axis and angle of rotation.

Although a qubit's state can be any point on the Bloch sphere, \emph{measuring} its value yields 0 or 1, each with some probability. Since $+$, $-$, $i$, and $-i$ lie on the equator, they are \emph{superpositions}, or combinations, of half $0$ and half $1$. Measuring them yields $0$ or $1$, each with a 50\% probability. The Measurement card, as well as kicking off after scoring a touchdown, corresponds to this: if the qubit is not at 0 or 1, then it must be at $+$, $-$, $i$, or $-i$ since those are the only other states used in the game. Measuring the qubit will yield $0$ or $1$ with a 50:50 chance. On the other hand, if the qubit is at 0, a measurement is certain to yield 0, so the state does not change, and if the qubit at $1$, a measurement is certain to yield $1$, so the state does not change.

%-------------------------------------------------------------------------------
% Section
%-------------------------------------------------------------------------------

\section{Research on Student Interest, Self-confidence, and Enjoyment}

To evaluate whether Qubit Touchdown can affect students' interest and self-confidence in learning quantum computing, as well as whether the game is easy to learn and fun to play, we conducted a study consisting of 107 public high school students across seven Precalculus, AP Statistics, and/or AP Physics~1 classes over four periods in one school day, which necessitated combining some classes in the library for the activity.

\begin{table*}
\caption{\label{table:survey}The survey questions given to students. Part 1 was given before students played the game. Part 2 was given after playing the game. Part 3 was given after explaining the game's connections to quantum computing. For all questions, respondents circled one of five options on the Likert scale: strongly disagree, disagree, neutral, agree, and strongly agree. Results are given as the percentage of students who answered at each Likert scale level. Totals may not equal 100\% due to rounding.}
\begin{ruledtabular}
\begin{tabular}{lp{0.5\linewidth}ccccc}
    \multicolumn{2}{l}{Question} & Strongly Disagree & Disagree & Neutral & Agree & Strongly Agree \\
    \hline 
    \multicolumn{2}{l}{Part 1} \\
    1. & I am interested in exploring quantum computing. & 13.1\% & 13.1\% & 37.4\% & 28.0\% & 8.4\% \\
    2. & I am confident in my ability to learn quantum computing. & 6.5\% & 22.4\% & 34.6\% &  29.9\% & 6.5\% \\
    \hline
    \multicolumn{2}{l}{Part 2} \\
    3. & After playing Qubit Touchdown, I am interested in exploring quantum computing. & 3.7\% & 21.5\% & 28.0\% & 39.3\% & 7.5\% \\
    4. & After playing Qubit Touchdown, I am confident in my ability to learn quantum computing. & 4.7\% & 10.3\% & 37.4\% & 40.2\% & 7.5\% \\
    5. & I enjoyed playing Qubit Touchdown. & 1.9\% & 1.9\% & 12.1\% & 46.7\% & 37.4\% \\
    6. & The rules of Qubit Touchdown were easy to learn. &  0.0\% & 0.0\% & 2.8\% & 31.8\% & 65.4\% \\
    \hline
    \multicolumn{2}{l}{Part 3} \\
    7. & After playing Qubit Touchdown and learning about its connections to concepts and rules in quantum computing, I am interested in exploring quantum computing. & 7.5\% & 15.0\% & 30.8\% & 37.4\% & 9.3\% \\
    8. & After playing Qubit Touchdown and learning about its connections to concepts and rules in quantum computing, I am confident in my ability to learn quantum computing. & 3.7\% & 4.7\% & 22.4\% & 54.2\% & 15.0\% \\
    9. & Playing Qubit Touchdown made it easier to learn quantum computing concepts. & 1.9\% & 0.0\% & 12.1\% & 48.6\% & 37.4\% \\
    10. & Playing Qubit Touchdown made quantum computing concepts more familiar. & 1.9\% & 1.9\% & 11.2\% & 58.9\% & 26.2\% \\
    11. & Playing Qubit Touchdown made quantum computing concepts less intimidating. & 2.8\% & 2.8\% & 15.9\% & 44.9\% & 33.6\% \\
    12. & Playing Qubit Touchdown provided a better learning experience for learning quantum computing concepts, compared to traditional instructional methods without games. & 0.9\% & 2.8\% & 11.2\% & 47.7\% & 37.4\% \\
    13. & I would recommend Qubit Touchdown to others as a learning tool for quantum computing. & 0.9\% & 0.9\% & 15.0\% & 43.9\% & 39.3\%
\end{tabular}
\end{ruledtabular}
\end{table*}

This study was approved by the Creighton University Institutional Review Board, record number 2004673-01, and as such, students were given a research information sheet and told that participation in the study was optional. They were told that they would be playing a quantum computing board game called Qubit Touchdown, and that if they wanted to participate in the study, they would be surveyed before playing the game, after playing the game, and then after learning about how the game connects to quantum computing. The survey questions for all three parts are shown in Table~\ref{table:survey}, and students responded to each question on a Likert scale of strongly disagree, disagree, neutral, agree, and strongly agree. Then, students were told the following definition of a quantum computer:
\begin{quote}
    A quantum computer is a new kind of computer that’s based on the laws of quantum physics. It can do certain things faster than normal computers because it follows a different set of rules.
\end{quote}
Following this, the survey was distributed, and students filled out Part 1, i.e., questions 1 and 2. After this, students were taught the rules of the game, similarly to the description in Section~\ref{sec:howtoplay}, so in analogy to football and without technical references. After students played the game with each other, they filled out Part 2 of the survey, i.e., questions 3 through 6. Finally, they were taught how Qubit Touchdown connects to quantum computing, similarly to Section~\ref{sec:connections}, and then they filled out Part 3 of the survey, i.e., questions 7 through 13. Results from the survey are also shown in Table~\ref{table:survey}.

\begin{table}
\caption{\label{table:survey_results}Overall agree and disagree percentages ($A$ and $D$) for the survey results.}
\begin{ruledtabular}
\begin{tabular}{ccc}
    Question \# & $A$ & $D$ \\ 
    \hline
    Q1 & 36.4\% & 26.2\% \\
    Q2 & 36.4\% & 28.9\%  \\
    Q3 & 46.8\% & 25.2\% \\
    Q4 & 47.7\% & 15.0\% \\
    Q5 & 84.1\% & 3.8\%  \\
    Q6 & 97.2\% & 0.0\% \\
    Q7 & 46.7\% & 22.5\%  \\
    Q8 & 69.2\% & 8.4\%  \\
    Q9 & 86.0\% & 1.9\%  \\
    Q10 & 85.1\% & 3.8\%  \\
    Q11 & 78.5\% & 5.6\%  \\
    Q12 & 85.1\% & 3.7\%  \\
    Q13 & 83.2\% & 1.8\% 
\end{tabular} 
\end{ruledtabular}
\end{table}

For a simple first analysis, we treat the Likert scale results as ordinal data and combined agree and strongly agree responses, and disagree and strongly disagree responses, to form overall agree and disagree values, $A$ and $D$, respectively, as shown in Table~\ref{table:survey_results} for all thirteen questions (Q1 to Q13), ignoring neutral responses \cite{Redish1998}. The results indicate that students were overall very favorable about their experience of playing Qubit Touchdown. For example, 78.5\% of the students reported that playing Qubit Touchdown made quantum computing concepts less intimidating (Q11), and 84.1\% of the students reported that they enjoyed playing the game (Q5). Similarly, 83.2\% of students recommended that Qubit Touchdown be used as a learning tool for quantum computing (Q13). Students' interest in exploring quantum computing at different time points (before playing the game, after playing, and after learning the game's connections to quantum computing) was assessed in Q1, Q3, and Q7, and we compare them by calculating the shift in overall agree percentages, presented as ``favorable shift'' values (movement toward agreement) in Table~\ref{table:results}. After playing the game, students' interest improved, but learning the game's connections to quantum computing did not further improve their interest. Similarly, students' confidence in their ability to learn quantum computing was assessed at the three time points in Q2, Q4, and Q8, and the favorable shifts are also presented in Table~\ref{table:results}. Their self-confidence improved after playing the game and again after learning the game's connection to quantum computing.

\begin{table}
\caption{\label{table:results}Shifts in favorable student responses between various questions pairs along with the statistical significance of the shift (Wilcoxon $p$-value). Favorable shift is the shift in $A$ values between the questions ($\Delta A$). Questions 1, 3, and 7 deal with student interest in exploring and learning about quantum computing while questions 2, 4, and 8 deal with student self-confidence in their ability to  learn quantum computing.}
\begin{ruledtabular}
\begin{tabular}{ccc}
    Question Pairs & Favorable Shift ($\Delta A$) & Wilcoxon $p$-value \\ 
    \hline
    Q1$\rightarrow$Q3 & 10.4\% & $p < 0.05$ \\
    Q3$\rightarrow$Q7 & -0.1\% & $p > 0.05$ \\
    Q1$\rightarrow$Q7 & 10.3 \% & $p < 0.05$ \\
    \hline
    Q2$\rightarrow$Q4 & 11.3\% & $p < 0.001$ \\
    Q4$\rightarrow$Q8 & 21.5\% & $p < 0.001$ \\
    Q2$\rightarrow$Q8 & 32.8\% & $p < 0.001$
\end{tabular}
\end{ruledtabular}
\end{table}

The simple binomial analysis of the Likert scale data above, however, has several disadvantages: it loses the information provided by neutral responses, it loses information from changes in responses from dis/agree to strongly dis/agree or vice-versa, and is not statistically rigorous. Because we have ordinal data which is non-normally distributed and our relevant questions came in sets of three, we first use a Friedman test to determine if the sets of responses to Q1, Q3, and Q7 (as well as Q2, Q4, and Q8) are statistically different when measured at the different time points. We found that there was a significant main effect of measuring student interest in exploring quantum computing at different time points (Q1, Q3, and Q7) by utilizing a Friedman test ($\chi^2 = 8.9589$, $d_f = 2$, $p < 0.05$). In other words, student interest in quantum computing is indeed different before versus after playing the game. In terms of exploring students' self-confidence in learning quantum computing, we found a stronger effect: there was a significant main effect of time point for confidence in their ability to learn quantum computing by utilizing a Friedman test ($\chi^2 = 49.004$, $d_f = 2$, $p < 0.001$).

Since the Friedman test results were significant and to help control for inflated Type I error due to multiple comparisons, we follow up the Friedman tests with pairwise paired Wilcoxon signed-rank tests with a continuity correction. All $p$-values were two-sided and considered significant if $p < 0.05$. The $p$-values from this analysis are reported in Table~\ref{table:results}. For students' interest in quantum computing, there was a statistically significant increase in student interest between Q1 and Q3 ($p < 0.05$) and between Q1 and Q7 ($p < 0.05$), but not between Q3 and Q7 ($p > 0.05$). For students' self-confidence, there was a statistically significant increase between Q2 and Q4 ($p < 0.001$), between Q2 and Q8 ($p < 0.001$), and between Q4 and Q8 ($p < 0.001$), and they are consistent with the favorable shifts. Table~\ref{table:results} gives both the simple favorable shift (movement towards more agreement) and the Wilcoxon $p$-values for question pairs (which shows that the shift is or is not statistically significant).

In summary, a statistical analysis of the Likert scale data shows that students seem to be both more interested in quantum computing and more self-confident in their ability to learn quantum computing as a result of playing Qubit Touchdown and listening to a short lecture on its connections to quantum computing.

%-------------------------------------------------------------------------------
% Section
%-------------------------------------------------------------------------------

\section{How to Obtain Qubit Touchdown}

\begin{figure}
    \includegraphics[width=3in]{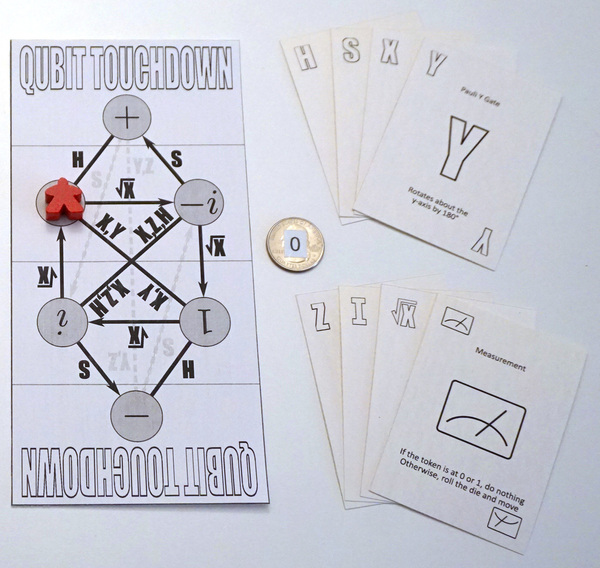}
    \caption{\label{fig:DIY}A do-it-yourself version of Qubit Tocuhdown.}
\end{figure}

One can make their own copy of Qubit Touchdown, an example of which is shown in Fig.~\ref{fig:DIY}, and video instructions are available.\cite{QubitTouchdown-DIY} For the game board, a black-and-white printable is available on the creator's website,\cite{QubitTouchdown-Website} or it could be hand-drawn. For the cards, black-and-white printables are also available on the creator's website, but they could also be hand-drawn on index cards, card stock, cardboard, or other materials. For the football, any object that can be moved from position to position on the game board will suffice. For the binary die, a coin or any other object to generate a random outcome can be used, including a standard six-sided die with even outcomes interpreted as $0$ and odd outcomes as $1$.

Another option is to order a professionally produced copy from The Game Crafter, a print-on-demand manufacturer in the United States.\cite{QubitTouchdown-TGC} The complete game in Fig.~\ref{fig:overview} was obtained this way, and it includes a game box.

%-------------------------------------------------------------------------------
% Section
%-------------------------------------------------------------------------------

\section{Conclusion}

Quantum computing is a critical area of science and technology, and games may play a role in developing the workforce for the field. Our study involving 107 public high school students indicates that playing Qubit Touchdown appeared to improve students' confidence in their ability to learn quantum computing by 11.3\%, and learning the game's connections to quantum computing yielded a net improvement in self-confidence of 32.8\%. Students' interest in the field, however, was modestly improved by 10.3\% by playing the game and learning its technical connections, and we speculate that it may be because the students were mostly juniors and seniors, who may already have plans for after they graduate from high school. These shifts in attitudes were found to be statistically significant. Students found the game enjoyable and easy to learn, and found that the game made it easier to learn quantum computing. They recommended it as a learning tool. As such, Qubit Touchdown could be a valuable resource for learning and outreach, giving high students students confidence to pursue technical fields they may have initially considered out of reach.

%-------------------------------------------------------------------------------
% Acknowledgments
%-------------------------------------------------------------------------------

\begin{acknowledgments}
    We thank Dr.~Jack Taylor in the Creighton University Biostatistical Core for help with the statistical analysis.  

    This material is based upon work supported in part by the National Science Foundation EPSCoR Cooperative Agreement OIA-2044049, Nebraska’s EQUATE collaboration. Any opinions, findings, and conclusions or recommendations expressed in this material are those of the author(s) and do not necessarily reflect the views of the National Science Foundation.
\end{acknowledgments}

%-------------------------------------------------------------------------------
% References
%-------------------------------------------------------------------------------

\bibliography{refs}

\end{document}